\documentclass[prd,aps,twocolumn,showpacs,preprintnumbers,amsmath,amssymb,superscriptaddress,nofootinbib]{revtex4-1}
\usepackage[dvips]{graphicx}
\usepackage{epsf}
\usepackage{amsmath}
\usepackage{amssymb}
\usepackage{amsfonts}
\usepackage{times}
\usepackage[usenames]{color}
\usepackage[normalem]{ulem}
\usepackage[T1]{fontenc}
\usepackage[dvipsnames]{xcolor}

\usepackage{graphicx}
\usepackage{dcolumn}
\usepackage{bm}
\usepackage{color}

\begin{document}

\title{An accurate method of modeling cluster scaling relations in modified gravity}

\author{Jian-hua He}
\email[Email address: ]{jianhua.he@durham.ac.uk}
\affiliation{INAF-Observatorio Astronomico, di Brera, Via Emilio Bianchi, 46, I-23807, Merate (LC), Italy}
\affiliation{Institute for Computational Cosmology, Department of Physics, Durham University, Durham DH1 3LE, United Kingdom}

\author{Baojiu Li}
\affiliation{Institute for Computational Cosmology, Department of Physics, Durham University, Durham DH1 3LE, United Kingdom}

\preprint{DURAST/2015/0005}

\begin{abstract}
We propose a new method to model cluster scaling relations in modified gravity. Using a suite of non-radiative hydrodynamical simulations, we show that the scaling relations of cumulative gas quantities, such as the Sunyaev Zel'dovich effect (Compton-y parameter) and the x-ray Compton-Y parameter, can be accurately predicted using the known results in the $\Lambda$CDM model with a precision of $\sim3\%$. This method provides a reliable way to analyze the gas physics in modified gravity using the less-demanding and much more efficient pure cold dark matter simulations. Our results therefore have important theoretical and practical implications in constraining gravity using cluster surveys.
\end{abstract}

\maketitle
\section{Introduction}
Clusters of galaxies are the largest gravitationally-bound objects in the Universe, and their formation and evolution are strongly affected by gravity. Therefore, their abundance and distribution are sensitive to the nature of gravity, making them an ideal cosmological probe to test and constrain gravity theories~\cite{DETF}. With them being observed in various ways, including x-ray, optical richness, the Sunyaev Zel'dovich (SZ) effect~\cite{SZ} and weak lensing, we are entering a new era of precision cluster cosmology. The upcoming eROSITA project~\cite{eROSITA}, for example, will probe up to $\sim10^5$ galaxy clusters on the whole sky out to redshift $z\geq1$, and there will be follow-ups and synergies with other surveys, such as Euclid~\cite{Euclid}, which will further improve the determination of cluster properties.

However, the robustness of the cosmological constraints from cluster surveys depends strongly on the accuracy of the measurement of cluster mass. The latter is highly non-trivial and, indeed, is one of the main challenges in cluster cosmology. Although the mass of a cluster can be directly derived from the gas density and temperature profiles in x-ray surveys, this method requires high-quality spectra and therefore long exposure time, which is expensive for distant clusters ($z>0.5$) and also for a large number of clusters. It also assumes hydrostatic equilibrium in clusters, which induces a systematic error of $\sim10\%$ in the cluster mass estimation~\cite{hydrostatic}.

In practice, this difficulty can be overcome by using scaling relations which relate the cluster mass to mass proxies that are combinations of observables which are easy to measure and have small intrinsic scatters. These scaling relations can be calibrated either by observations or by hydrodynamical simulations~\cite{Fabjan,cluster_observation}. In observations, this can be done by using either complementary observables, e.g., weak lensing, or a subset of observational data with better mass determination. In the $\Lambda$CDM model, with ever-improving resolution and modelling of baryonic physics, simulation calibration of the scaling relations are becoming more accurate and found to have a good match with observations~\cite{cluster_observation}.

However, using cluster observations to test gravity in a fully self-consistent way is nontrivial. People usually compare the observational data with theoretical predictions for certain representative modified gravity models~\cite{MGcluster}. One important point, which is sometimes missed, is that the calibration of the observational data itself in such models are generally much more nontrivial. The intrinsic non-linearity in these models not only changes the various scaling relations from the known results in $\Lambda$CDM, but also their intrinsic scatters and correlations with other cluster properties. For example, in the widely-studied $f(R)$ gravity model~\cite{HuS}, the lensing and hydrostatic masses of a cluster can be different, and the difference depends on the screening effect which further depends on many factors such as the cluster's redshift, mass, and environment(see Ref.~\cite{frscreening} for detailed discussion). This makes it difficult, both in observations and simulations, to calibrate the scaling relations that involve hydrostatic observables, such as the x-ray luminosity and temperature.

To tackle this important but so far not well-understood issue, in this work, we propose a new method to model cluster scaling relations in modified gravity. We show that the effect of modified gravity can be modelled as a rescaling of the cluster gas mass fraction. This rescaling can be quantified directly using less demanding pure dark matter simulations, and then used to accurately predict cluster scaling relations in modified gravity, based on the known results in $\Lambda$CDM. We demonstrate the accuracy of this method by comparing the predictions from our rescaling method for three mass proxies -- the x-ray luminosity $L_{\rm X}$, the SZ Compton-y parameter $Y_{\rm SZ}$ and the x-ray Compton y parameter $Y_{\rm X}$~\cite{Y_X} --with those from non-radiative hydrodynamical simulations. We find that the method works a $\sim3\%$ accuracy for the latter two proxies.

This paper is organised as follows: In Sec.~\ref{simulation}, we introduce the hydrodynamical simulations used in this work. In Sec.~\ref{gastracer}, we introduce the halo catalogs constructed in this work. In Sec.~\ref{profile}, we describe the features of profiles of the gaseous halos in our simulations. In Sec.~\ref{fraction}, we compare the gas fractions in $\Lambda$CDM and $f(R)$ gravity. In Sec.~\ref{scal}, we show that the effect of modified gravity can be modelled as a rescaling of the cluster gas mass fraction. In Sec.~\ref{conclusions}, we summarise and conclude this work.

\section{ Hydrodynamical simulations\label{simulation}}
We ran a suite of hydrodynamical simulations using {\sc ecosmog} \cite{ECOSMOG}, a modified gravity simulation code based on the publicly available {\sc ramses} \cite{RAMSES} code. We simulated an $f(R)$ model which exactly reproduces the $\Lambda$CDM background expansion history~\cite{frmodel}, using a box of size $L_{\rm box}=100 h^{-1}{\rm Mpc}$ that contains $N = 256^3$ cold dark matter particles. The cosmological parameters are the same as the Planck \cite{planck} best-fit $\Lambda$CDM model ($\Omega_b^0=0.045, \Omega_c^0=0.271, \Omega_d^0=0.684, h=0.671, n_s=0.962$, and $\sigma_8=0.834$). Initial conditions were generated using the {\sc mpgrafic} package~\cite{inicon} at $z=49$.

Due to the expensive cost of hydrodynamic simulations, we focus only on the $f(R)$ model with $f_{R0}=-10^{-5}$ (where $f_{R0}$ is the present value of ${\rm d}f/{\rm d}R$). We run three realizations in total for the $f(R)$ model, and for each $f(R)$ simulation, we run two $\Lambda$CDM simulations as control, one with exactly the same initial conditions as the corresponding $f(R)$ run and the other with a rescaled baryon fraction, $\frac{3}{4}\Omega_b^0$, while keeping $\Omega_m^0 = \Omega_b^0 + \Omega_c^0$ unchanged.

We assume that all the baryon component in our simulation is non-radiative ideal gas which obeys $P_{\rm gas}=k_BT_{\rm gas}\rho_{\rm gas}/(\mu m_p)\,,$
where $P_{\rm gas}$ is the thermal pressure of gas, $\rho_{\rm gas}= \mu m_p n_{\rm gas}$ is the density, $k_B$ the Boltzmann constant, $\mu$ the mean molecular weight, $n_{\rm gas}$ the number density of gas particles and $m_p$ the proton mass. We also assume that the gas is fully ionised with compositions of electrons, hydrogen and helium. We take the primordial mass fraction of hydrogen as $n_{\rm H}/(n_{\rm H}+4n_{\rm {He}})\approx 0.75$, so that the mean molecular weight is $\mu\approx0.59\,.$ Finally, for a monatomic gas the adiabatic index is
$\gamma =\frac{5}{3}$.

\section{ Gas tracer particles and effective halo catalogs\label{gastracer}} 
\begin{figure}
\includegraphics[width=3in,height=2.8in]{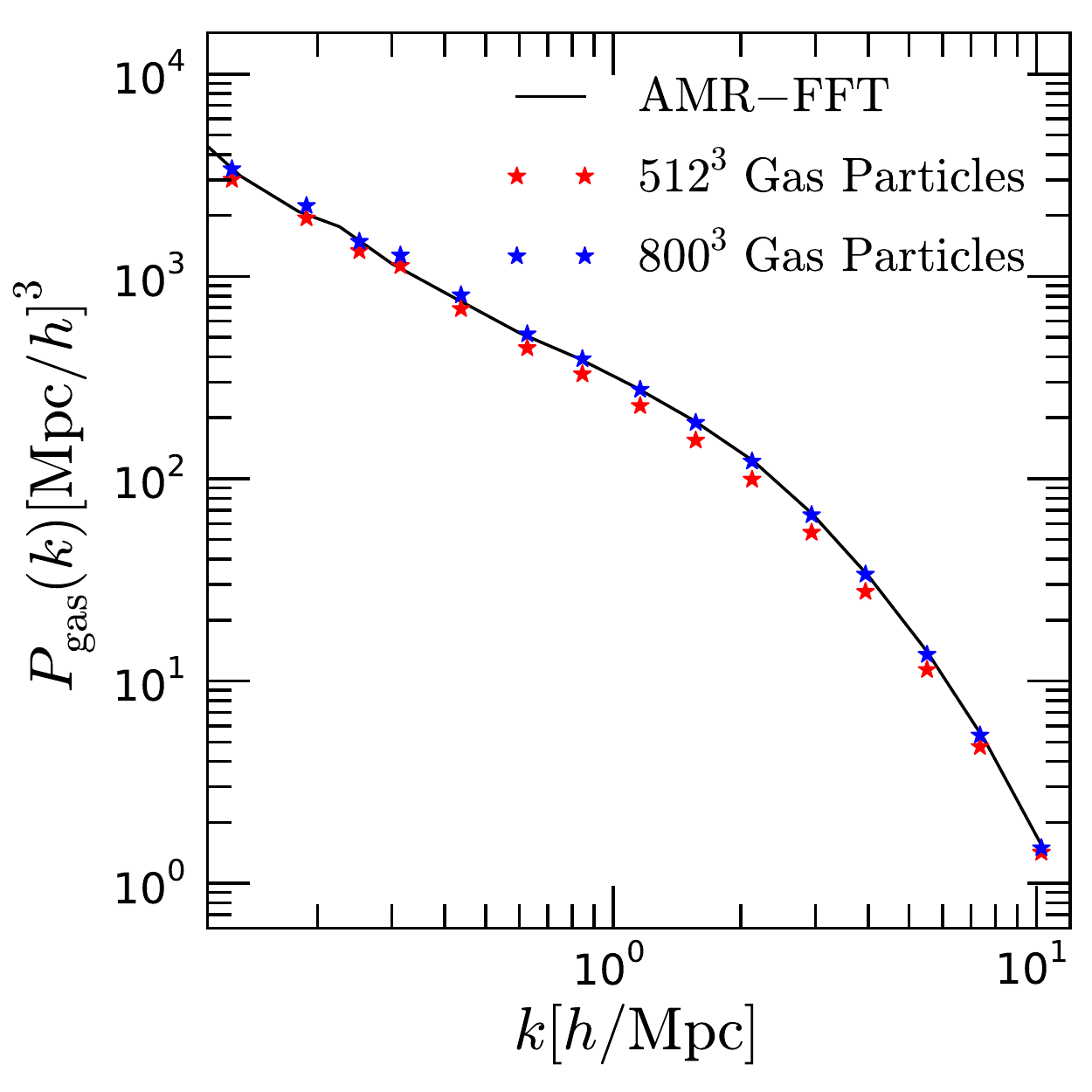}
\caption{A comparison of the power spectra from the gas density field on AMR grids (black solid line) and the gas tracer particles with two different numbers of particles (stars).   With a large enough number of particles (e.g. $N_{\rm gas}=800^3$ ), the tracer particles and the original gas density field have almost the same power spectra, which indicates that the tracer particles can accurately represent the original gas density field on the AMR grids.} \label{tracer}
\end{figure}

{\sc ramses} is a mesh-based code which does not define gas particles by default. In order to trace the gas density field on the simulation grids, we sample it with tracer particles following a Poisson process. The tracer particles have a uniform mass and their local mean is set to be proportional to the local gas density on the grid. By comparing the density power spectra of the tracer particles and of the density field on the grids, we find that with a large enough number of particles (e.g. $N_{\rm gas}=800^3$ as shown in Fig.~\ref{tracer}), the tracer particles can accurately represent the original gas density field on the simulation grids.

Our halo catalogs are constructed using a modified version of the publicly available {\sc Amiga} Halo Finder ({\sc Ahf})~\cite{AHF}. We call the halo catalog produced using the true density field, the {\it standard catalog}. On the other hand, as discussed in \cite{EFFhalos}, the modified Poisson equation in $f(R)$ gravity can be cast into the standard form
\begin{equation}
\nabla \phi=4\pi G a^2 \delta \rho_{\rm eff}\quad,
\end{equation}
by defining the effective energy density $\delta\rho_{\rm eff}$ which incorporates all the modified gravity effect, with $G$ being Newton's constant. The halo catalog constructed using 
$\delta \rho_{\rm eff}$ is referred to as {\it effective catalog} (see \cite{freffpk,EFFhalos} for technical details).
\begin{figure}
\includegraphics[width=3.5in,height=5.0in]{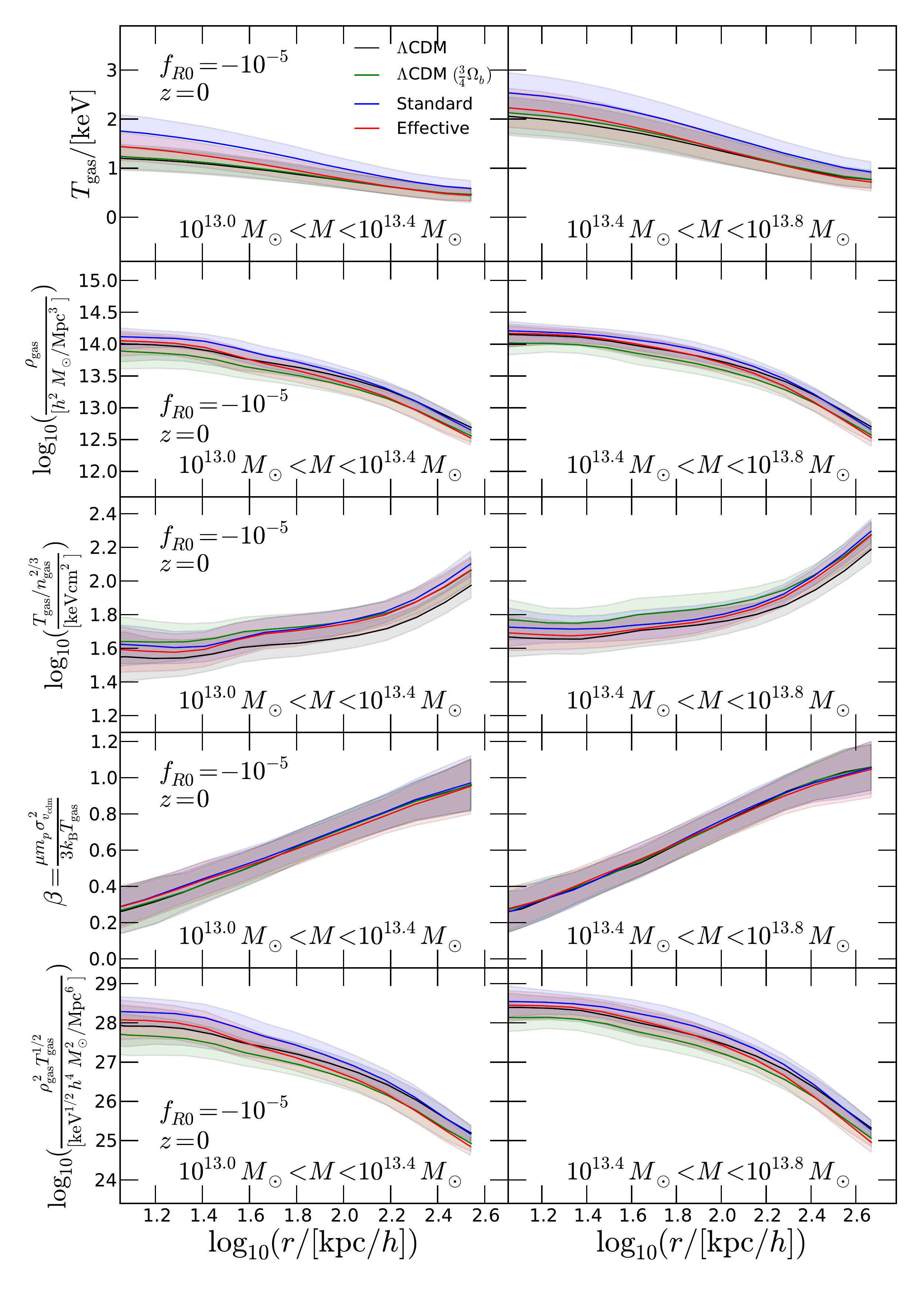}
\caption{Profiles for different gas quantities from four different halo catalogs: $\Lambda$CDM (black lines), $\Lambda$CDM with rescaled baryon fraction $\frac{3}{4}\Omega_b$ (green lines), standard halo catalog in $f(R)$ gravity (blue lines), and effective halo catalog in $f(R)$ gravity (red lines).
From top to bottom we have the profiles of gas temperature $T_{\rm gas}$, density $\rho_{\rm gas}$, entropy $S_{\rm gas}=T_{\rm gas}/n_{\rm gas}^{2/3}$, $\beta=\mu m_p \sigma_{v_{\rm cdm}}^2/(3k_B T)$ and surface brightness $\rho_{\rm gas}^2T_{\rm gas}^{1/2}$.
The two columns in each row are halos within different mass bins.
The shaded regions represent the $1\sigma$ scatter.
}\label{figone}
\end{figure}

\section{ Profiles\label{profile}} We use halos in the mass range of $10^{13}M_{\odot}/h<M<10^{13.8}M_{\odot}/h$, in which most halos are unscreened for $ f_{R0}=-10^{-5}$. This range is further divided into two different bins: $10^{13}M_{\odot}/h<M<10^{13.4}M_{\odot}/h$ and $10^{13.4}M_{\odot}/h<M<10^{13.8}M_{\odot}/h$. We consider profiles only within $10 {\rm kpc}/h<r<R_{\rm vir}$, where the virial radius $R_{\rm vir}$ is defined by an overdensity of $300$ with respect to the critical density.

The gas temperature profiles are shown in the first row of Fig.~\ref{figone}. We can see that the effective halos in $f(R)$ gravity and the $\Lambda$CDM halos are very similar to each other. This is as expected since gas temperature is closely related to the gravitational potential of halos. The physical reason behind this is that, during gas accretion, the shocked gas converts the energy that it gains from gravitational infall into the thermal energy of itself. Since the bulk motion of gas accounts for only a small fraction of the total gained energy, the gravitational potential energy is mainly converted into the its thermal energy, and so the temperature of the gas is mainly determined by the gravitational potential.

Compared with the temperature profile, the density profile is more complicated. Unlike cold dark matter, gas has a core at the halo center and its density profile is flat at $r\lesssim R_{\rm vir}/20$. The specific entropy of gas does not vary significantly in this core region, as shown in the third row of Fig.~\ref{figone}, and the shocked gas can be considered as adiabatic~\cite{Gasprofiles}. From simulations, we find that effective $f(R)$ halos and $\Lambda$CDM halos have very close density and entropy profiles in this core region (see the second and third rows in Fig.~\ref{figone}).

The shape of the gas density profile outside the core region ($r>R_{\rm vir}/2$) can be described by the $\beta$ model~\cite{Gasprofiles}, where $\beta$ is the ratio between the specific kinetic energy $\mathcal{K}_{\rm cdm}$ (kinetic energy per unit mass) of cold dark matter and the specific internal energy $\mathcal{E}_{\rm gas}$ (internal energy per unit gas mass) of gas
\begin{equation}
\beta \equiv \frac{\mathcal{K}_{\rm cdm}}{\mathcal{E}_{\rm gas}}=\frac{\sigma_{v_{\rm cdm}}^2/2}{P_{\rm gas}/\rho_{\rm gas}/(\gamma -1)}=\frac{\mu m_p \sigma_{v_{\rm cdm}}^2}{3k_B T}\quad.
\end{equation}
If we consider the 3-D velocity dispersion $\sigma_{v_{\rm cdm}}^2$ as a measure of the temperature of cold dark matter~\cite{halotem}, then $\beta$ is simply the ratio of temperatures between cold dark matter and gas.

As shown in Fig.~\ref{figone}, the temperature profile does not vary significantly outside the core region and so gas and cold dark matter can be roughly treated as isothermal. Furthermore, if we assume a hydrostatic equilibrium between them, the shapes of the gas and dark matter density profile are related by~\cite{book}
\begin{equation}
\beta \approx \frac{d\ln \rho_{\rm gas}/d\ln r}{d\ln \rho_{\rm cdm}/d \ln r}\quad.\label{betam}
\end{equation}
In the $\Lambda$CDM model, this feature has been verified by a number of numerical simulations~\cite{betamodel}. As shown in the fourth row of Fig.~\ref{figone}, it is interesting to find that $\beta$ is almost identical in $f(R)$ and $\Lambda$CDM models. According to Eq.~(\ref{betam}),  outside the core region, the gas density profile in $f(R)$ gravity should trace the dark matter density profile in the same way as that in $\Lambda$CDM. As indicated in the second row of Fig.~\ref{figone}, at  $r>R_{\rm vir}/2$, in the standard catalog, the gas density profile in $f(R)$ gravity matches the $\Lambda$CDM result (blue and black lines); in the effective catalog, the gas density profile in the $f(R)$ model matches the $\Lambda$CDM one with a rescaled baryon fraction (red and green lines).

According to the gas density and temperature profiles, it can be expected that the profiles of other physical quantities that depend on gas density and temperature should have the following features: inside the core region, effective halos should behave similarly to $\Lambda$CDM halos; outside the core region, effective halos should resemble $\Lambda$CDM halos with rescaled gas fractions by $M^{f(R)}/M^{f(R)}_{\rm Eff}$. For illustrative purposes, we show the profile of the surface brightness $I(r)$. Since $I(r)$ is related to $\rho_{\rm gas}(r)^2T_{\rm gas}(r)^{1/2}$, 
in the last row of Fig.~\ref{figone} we only show the profile of $\rho_{\rm gas}(r)^2T_{\rm gas}(r)^{1/2}$. It is evident that the $\rho_{\rm gas}(r)^2T_{\rm gas}(r)^{1/2}$ profile has the features as just described the above.
\begin{figure*}
\includegraphics[width=6in,height=3.2in]{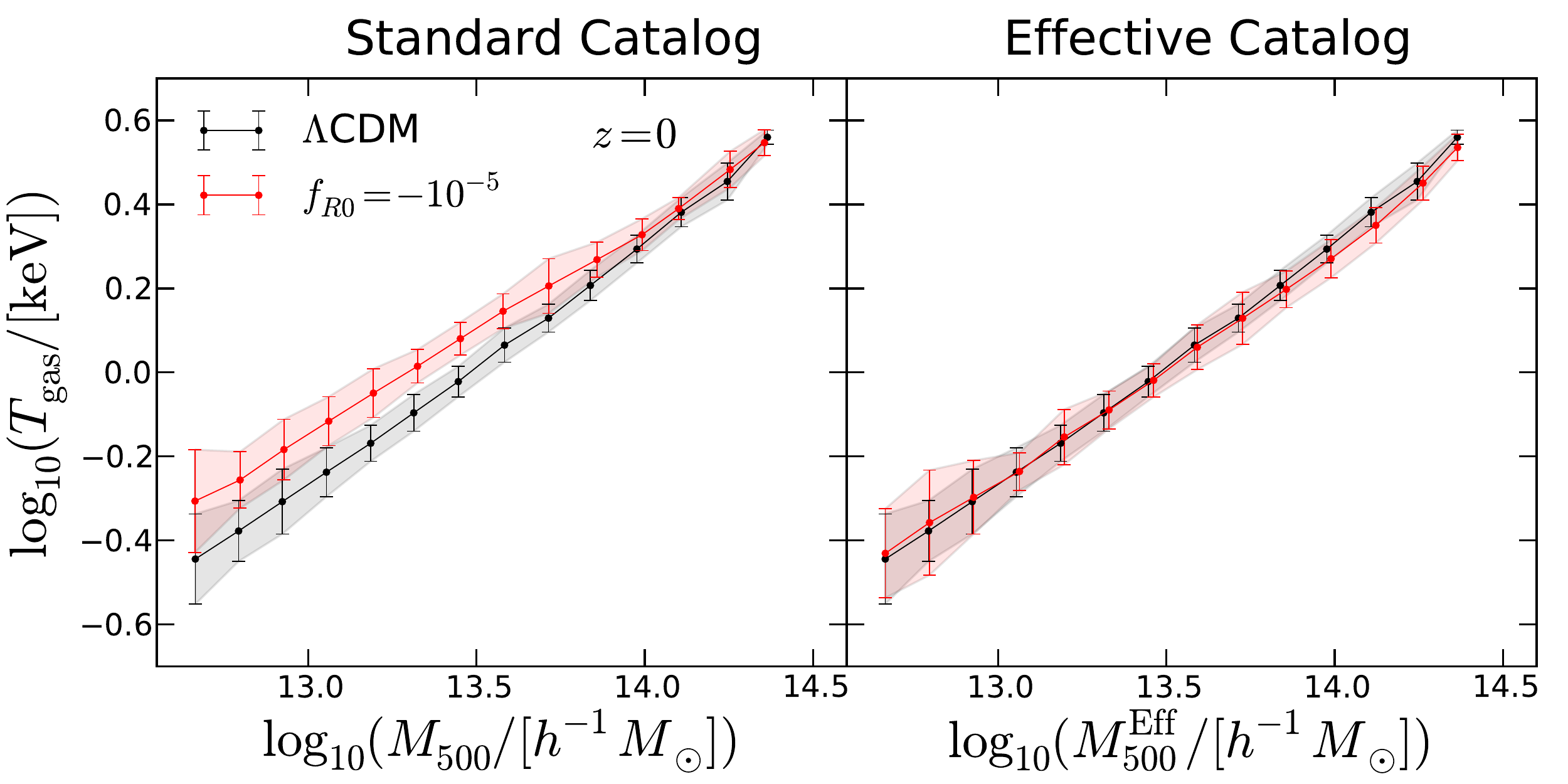}
\caption{Gas temperature as a function of halo mass in the $\Lambda$CDM model (black symbols) and $f(R)$ gravity (red symbols). Left: the standard halo catalogue of $f(R)$ gravity is used. Right: the effective halo catalogue of $f(R)$ gravity is used. The shaded regions represent $1\sigma$ scatter.
}\label{figtwo}
\end{figure*}
\section{gas fractions \label{fraction}}
\begin{figure*}
\includegraphics[width=6in,height=3.2in]{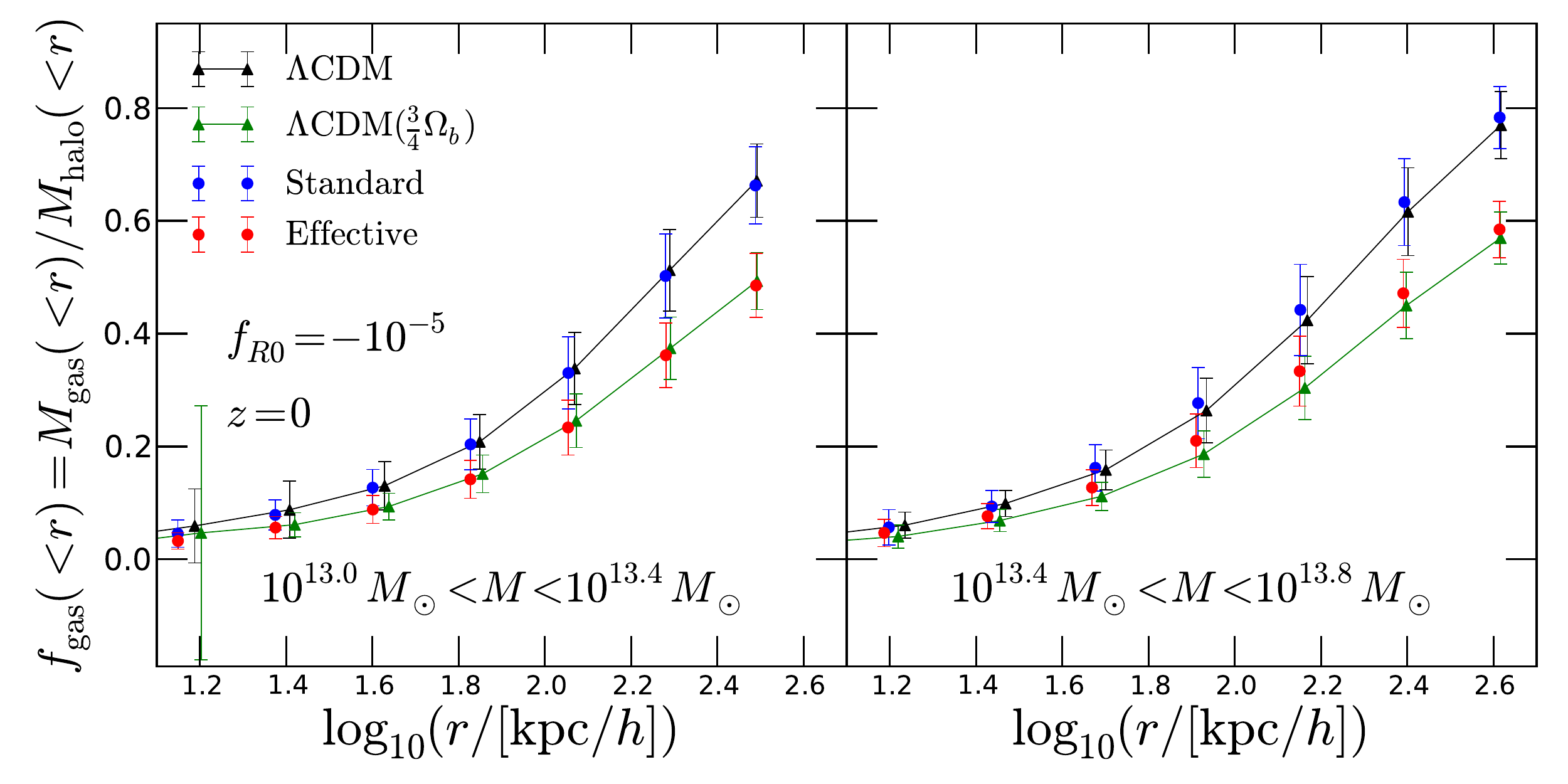}
\caption{The cumulative halo gas fraction as a
function of the distance from halo centre.  In
the standard $f(R)$ halo catalog (blue points), the gas fraction agrees very well with that of the $\Lambda$CDM halos (black
triangles). In contrast, the gas fraction of effective $f(R)$ halos (red points) agrees with that of $\Lambda$CDM
halos with rescaled $\Omega_b$ (green triangles). The left panel is for a mass bin in which
most halos are unscreened and the right panel is for a mass bin in which halos are partially screened.}\label{fractiongas}
\end{figure*}
After studying the gas density profiles, we turn to the cumulative halo gas mass fraction.
Since outside the core regions, the gas density profile in $f(R)$ gravity traces the dark matter density profile, 
the cluster gas fraction is a tracer of the true, rather than the effective, halo mass. 
In objects as large as galaxy clusters, it is well known that the gas fraction is more sensitive to $\Omega_b/\Omega_m$ rather than the theory of gravity. This is illustrated in Fig.~\ref{fractiongas} (also see Ref.~\cite{gasfrac}), where we can see that the standard halos in $f(R)$
gravity and $\Lambda$CDM halos have essentially identical gas fraction profiles (blue points versus black triangles).
On the other hand, the gas fraction in the $f(R)$ effective halo agrees with that of $\Lambda$CDM halos with rescaled $\Omega_b$.

\section{ Scaling relations\label{scal}} We next turn to the scaling relations of cumulative physical quantities. The cumulative quantities, such as the luminosity $L_{\rm X}$, describe the properties of clusters as a whole. In general, they are less sensitive to the profiles in the core region because the latter has a small volume and contributes little to the total cumulative value, but they are more sensitive to the profiles at a relatively larger radius.  As noted earlier, outside the core region, the profiles of effective halos in $f(R)$ gravity closely resemble those in the $\Lambda$CDM model with rescaled gas fractions.
This feature, as we shall show later, can help us build a connection between the scaling relations of physical quantities of $f(R)$ effective and $\Lambda$CDM halos.

\begin{figure*}
\includegraphics[width=5.5in,height=5in]{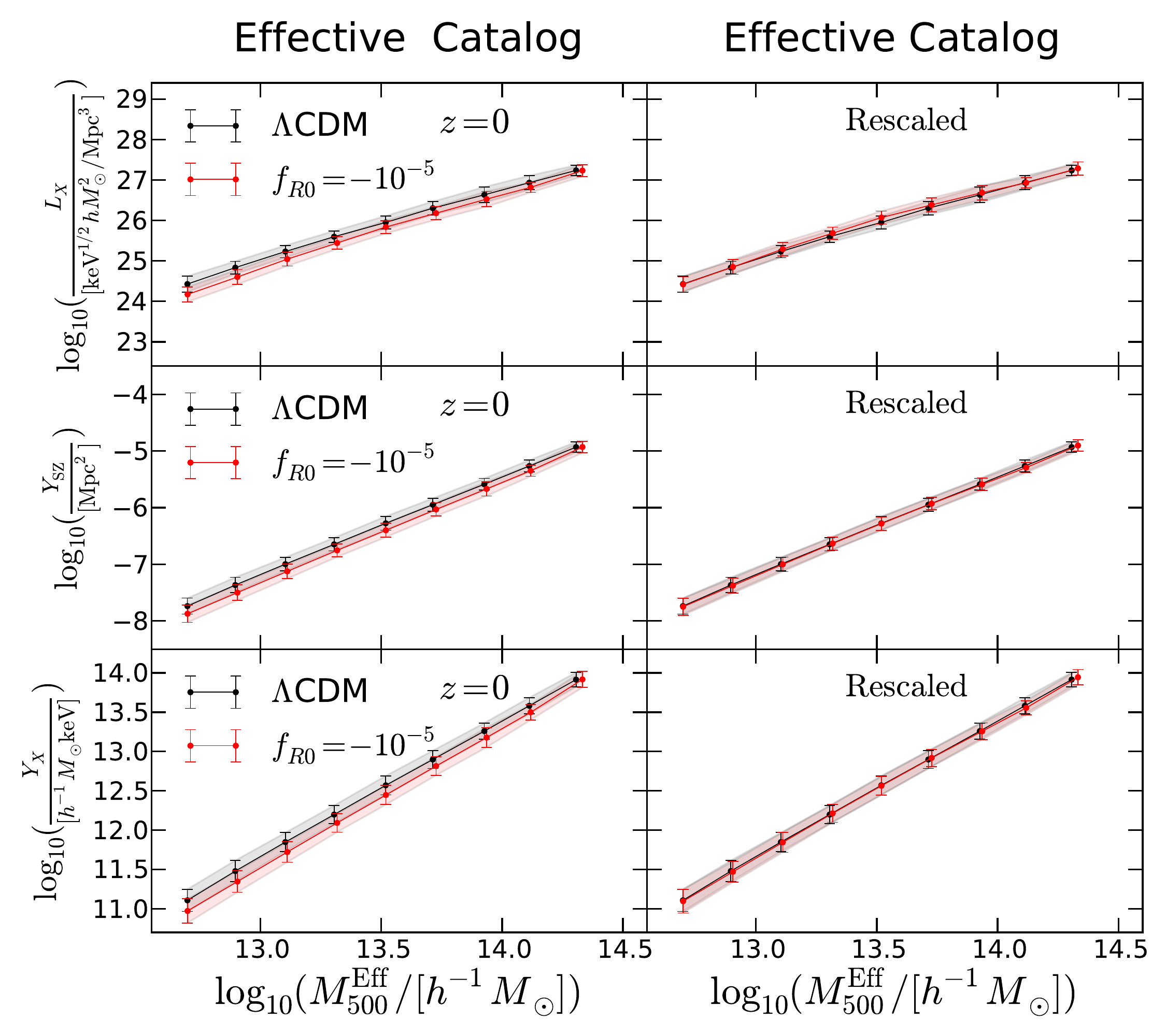}
\caption{Cluster x-ray luminosity (upper panels), $Y_{\rm SZ}$ parameter (middle panels) and $Y_{\rm X}$ parameter (lower panels) as a function of halo mass in the $\Lambda$CDM model (black symbols) and $f(R)$ gravity (red symbols). Left: results without the rescaling of Eqs.~(\ref{rescaling}). Right: results with the rescaling of Eqs.~(\ref{rescaling}). The results are almost identical in the two models in the latter case while differences are apparent in the former case. The shaded regions represent $1\sigma$ scatter. }\label{figthree}
\end{figure*}

We start with the temperature-mass relation. As shown in the left panel of Fig.~\ref{figtwo}, in the standard catalog, the gas temperature of unscreened $f(R)$ halos are higher than those of the $\Lambda$CDM halos with the same masses, which is consistent with what was found in~\cite{fr_volker} . However, in the effective catalog, as shown in the right panel of Fig.~\ref{figthree}, the $T_{\rm gas}$-$M$ relation in 
the two models agree very well with each other
\begin{equation}
T_{\rm gas}^{f(R)}(M^{f(R)}_{\rm Eff})=T_{\rm gas}^{\Lambda \rm CDM}(M^{\Lambda{\rm CDM}})\,.\label{TM}
\end{equation}
This is a natural result from the gas temperature profile as discussed above.

For gas densities, we have seen above that outside the core region we have
\begin{equation}
\rho_{\rm gas}^{f(R)}(r) \approx \frac{M^{f(R)}}{M^{f(R)}_{\rm Eff}}\rho_{\rm gas}^{\Lambda \rm CDM}(r)\propto \frac{M^{f(R)}}{M^{f(R)}_{\rm Eff}}\frac{\Omega_b}{\Omega_m}(r^2+r^2_{\rm core})^{-3\beta/2}\quad,\nonumber
\end{equation}
where $r_{\rm core}$ is the core radius. For an $f(R)$ effective halo with the same mass as a $\Lambda$CDM halo, $M^{f(R)}_{\rm Eff}=M^{\Lambda{\rm CDM}}$, from the above equation and the fact that $f(R)$ effective and $\Lambda$CDM halos have very similar temperature profiles, it follows that
\begin{equation}
\begin{split}
&\int_0^r dr 4\pi r^2 (\rho^{f(R)}_{\rm gas})^a (T^{f(R)}_{\rm gas})^b\\
\approx &\left(\frac{M^{f(R)}}{M^{f(R)}_{\rm Eff}}\right)^a\int_0^r dr 4\pi r^2 (\rho^{\Lambda \rm CDM}_{\rm gas})^a (T^{\Lambda \rm CDM}_{\rm gas})^b\quad,\label{scaling}
\end{split}
\end{equation}
where $a$ and $b$ are indices of power.

The above relation is one of the key results in this paper and it indicates the important relation of cumulative gas quantities between $f(R)$ effective halos and $\Lambda$CDM halos.
In order to test the validity and illustrate the use of Eq.~(\ref{scaling}), we investigate three most important and frequently used quantities in cluster surveys as examples.
The first one is the x-ray luminosity $L_{\rm X}$ which, for a cluster, can be written as
\begin{equation}
L_{\rm X}(<r)=\int_0^r dr 4\pi r^2 \rho_{\rm gas}^2 T_{\rm gas}^{1/2}\quad.
\end{equation}
$L_{\rm X}$ is sensitive to the details of the gas distribution in the central region and depends on the dynamical state of the cluster. As a cluster mass proxy, $L_{\rm X}$-$M$ has large scatters.

The second one is the integrated SZ Compton y-parameter
\begin{equation}
Y_{\rm SZ}(<r)=\frac{\sigma_T}{m_ec^2}\int_0^{r}dr4\pi r^2P_e\quad,
\end{equation}
where $\sigma_T$ is Stefan-Boltzmann constant, $m_e$ the electron mass and $c$ the speed of light. $P_e$ is the electron pressure, which is given by
$P_e = \frac{2+\mu}{5}n_{\rm gas}k_B T_{\rm gas}\,.$

The third one is the x-ray equivalent of the integrated SZ flux, the $Y_X$ parameter~\cite{Y_X}.
\begin{equation}
Y_{\rm X}(<r) = \bar{T}_{\rm gas}\int_0^r dr 4\pi r^2 \rho_{\rm gas}\quad,
\end{equation}
where $\bar{T}_{\rm gas}$ is the average mass-weighted temperature. $Y_{\rm X}$-$M$ relation is relatively insensitive to the dynamical state of clusters and to the detailed modeling of gas physics~\cite{Fabjan} and the $Y_{\rm X}$-$M$ scaling relation is often practically used in x-ray surveys (e.g.~\cite{Xray}).

From Eq.~(\ref{scaling}), it follows that
\begin{equation}
\begin{split}
\frac{M^{f(R)}_{\rm Eff}}{M^{f(R)}}&Y_{\rm SZ}^{f(R)}(M^{f(R)}_{\rm Eff})\approx Y_{\rm SZ}^{\Lambda \rm CDM}(M^{\Lambda{\rm CDM}}=M^{f(R)}_{\rm Eff})\,,\\
\left( \frac{M^{f(R)}_{\rm Eff}}{M^{f(R)}}\right)^2&L_{\rm X}^{f(R)}(M^{f(R)}_{\rm Eff})\approx L_{\rm X}^{^{\Lambda \rm CDM}}(M^{\Lambda{\rm CDM}}=M^{f(R)}_{\rm Eff})\,,\\
\frac{M^{f(R)}_{\rm Eff}}{M^{f(R)}}&Y_{\rm X}^{f(R)}(M^{f(R)}_{\rm Eff})\approx Y_{\rm X}^{\Lambda \rm CDM}(M^{\Lambda{\rm CDM}}=M^{f(R)}_{\rm Eff})\,.\label{rescaling}
\end{split}
\end{equation}

The numerical results of the scaling relations are shown in Fig.~\ref{figthree}, from which we can see that without the rescaling of Eqs.~(\ref{rescaling}), the $f(R)$ and $\Lambda$CDM models have very different scaling relations (left panels), but after the rescaling not only the mean value but also the scatters of the scaling relations in the two models highly resemble each other (right panels). In particular, for $Y_{\rm SZ}$ and $Y_{\rm X}$ after the rescaling the average relative difference between the models are only at a level of $3\%$ (see~Table.\ref{RelDiff}). Therefore, using this rescaling method, the complicated effect of modified gravity on the cluster scaling relations can be accurately modelled and the error is much smaller than the typical uncertainty caused by the modelling of baryonic physics in galaxy formation (e.g.~\cite{Fabjan}). 
\begin{table}
\caption{The average relative differences of scaling relations between $f(R)$ and $\Lambda$CDM models. }\label{RelDiff}
\begin{tabular}{|c|c|c|}
  \hline
& ${\rm Before \, Rescaling}$ & ${\rm After \, Rescaling}$  \\
\hline
  $<L_{\rm X}^{f(R)}/L_{\rm X}^{\Lambda \rm CDM}>-1$ & $27.2\%$ & $13.1\%$  \\
  $<Y_{\rm SZ}^{f(R)}/Y_{\rm SZ}^{\Lambda \rm CDM}>-1$ & $19.7\%$  & $3.1\%$  \\
  $<Y_{\rm X}^{f(R)}/Y_{\rm X}^{\Lambda \rm CDM}>-1$ & $19.6\%$  & $3.2\%$  \\
  \hline
\end{tabular}
\end{table}

\section{Discussion\label{conclusions}}
In this paper, we investigated the gas properties in a representative non-standard gravity model, $f(R)$ gravity, based on a suite of non-radiative hydrodynamical simulations. We studied both the profiles and the cluster scaling relations of gas properties. We found that the effective halos, which are proposed in~\cite{EFFhalos}, in $f(R)$ gravity have similar temperature profiles as $\Lambda$CDM halos with the same masses. For the gas density profile, effective halos closely resemble $\Lambda$CDM halos in the core region, while outside the core region they behave like $\Lambda$CDM halos with rescaled gas fractions.

Based on those observations, we have demonstrated that not only the mean value but also the scatters of the scaling relations of cumulative gas quantities in $f(R)$ effective halos with rescaled gas mass fractions are very similar to those in the $\Lambda$CDM model, cf.~Eq.~(\ref{scaling}). In particular, for $Y_{\rm SZ}$ and $Y_{\rm X}$, which are the two most frequently used mass proxies in cluster surveys, our rescaling method enables us to model the effect of modified gravity on the scaling relations with an accuracy of $\sim3\%$, which is much smaller than the typical uncertainty caused by the modelling of baryonic physics in galaxy formation (e.g.~\cite{Fabjan}). The error in modelling the impact of modified gravity on the scaling relations is a minor source of the overall uncertainty in cluster cosmology.
This, therefore, provides an accurate way to calibrate the scaling relations in $f(R)$ gravity and a way to correctly interpret them in the context of non-standard gravity theories. It will enable us to test gravity using cluster observations in a fully self-consistent way, and avoid substantial systematic biases. Moreover, this provides a reliable way to analyze the gas physics in modified gravity using much more efficient pure cold dark matter simulations with the known knowledge in the $\Lambda$CDM model.

Although we illustrated our idea using a specific $f(R)$ model with a fixed value of $f_{R0}$, our main results are expected to apply to other values of $f_{R0}$ and other $f(R)$ models as well. This is because the $f(R)$ model with the value of $f_{R0}$ chose here contains the totally screened, partially screened, and totally unscreened halos. As shown in Fig.~\ref{figthree}, our rescaling method works very well for all these halos that have different levels of screening. The difference in different models of $f(R)$ gravity and different values of $f_{R0}$ only lies in the exact mass ranges within which halos are totally unscreened, totally screened or partially screened. Therefore,  varying $f_{R0}$ does not affect the overall workability of our method and our method works for other $f(R)$ models as well.

From the above argument, our main results are also expected to apply to other models that employ the chameleon screening mechanism~\cite{Khoury} as well as dilaton~\cite{dilaton}, symmetron~\cite{symmetron} models and their generalisations~\cite{lisymtron}. Therefore, our method can have much wider applications and provide a useful way to constrain modified gravity theories using observations from upcoming cluster surveys such as eROSITA~\cite{eROSITA}.

{\bf Acknowledgments} J.H.H. acknowledges support of the Italian Space Agency (ASI), via contract agreement I/023/12/0. B.L. acknowledges support by the UK STFC Consolidated Grant ST/L00075X/1 and RF040335. This work has used the DiRAC Data Centric system at Durham University, operated by the Institute for Computational Cosmology on behalf of the STFC DiRAC HPC Facility (www.dirac.ac.uk). This equipment was funded by BIS National E-infrastructure capital grant ST/K00042X/1, STFC capital grant ST/H008519/1, STFC DiRAC Operations grant ST/K003267/1 and Durham University. DiRAC is part of the National E-Infrastructure. For
access to the simulation data and halo catalogs, please contact J.H.H.

\end{document}